\documentclass[cameraready]{Interspeech}
\title{NLE: Non-autoregressive LLM-based ASR by Transcript Editing}

\author{Avihu}{Dekel}
\author{Samuel}{Thomas}
\author{Takashi}{Fukada}
\author{George}{Saon}
\address{IBM Research}
\email{avihu.dekel@ibm.com}


\newcommand{\blue}[1]{\textcolor{blue}{#1}}
\newcommand{\red}[1]{\textcolor{red}{#1}}
\newcommand{\orange}[1]{\textcolor{orange}{#1}}
\usepackage{comment}
\usepackage{booktabs}
\usepackage{siunitx,xcolor,colortbl}
\newcommand{\method}{NLE}
\begin{document}

\maketitle

\begin{abstract}
While autoregressive (AR) LLM-based ASR systems achieve strong accuracy, their sequential decoding limits parallelism and incurs high latency.
We propose NLE, a non-autoregressive (NAR) approach that formulates speech recognition as conditional transcript editing, enabling fully parallel prediction.
NLE extracts acoustic embeddings and an initial hypothesis from a pretrained speech encoder, then refines the hypothesis using a bidirectional LLM editor trained with a latent alignment objective.
An interleaved padding strategy exploits the identity mapping bias of Transformers, allowing the model to focus on corrections rather than full reconstruction.
On the Open ASR leaderboard, NLE++ achieves 5.67\% average WER with an RTFx (inverse real-time factor) of 1630.
In single-utterance scenarios, NLE achieves 27x speedup over the AR baseline, making it suitable for real-time applications.
\end{abstract}

\section{Introduction}
\label{sec:intro}
LLM-based ASR systems~\cite{tang2024salmonn,saon2025granite} improve transcription accuracy by pairing a pretrained speech encoder with a pretrained LLM, typically via a learned projector that maps acoustic representations into the LLM embedding space.
In most existing systems, the LLM serves as an autoregressive decoder, generating text one token at a time.
While this design yields strong accuracy, the sequential nature of autoregressive decoding limits parallelism, incurs substantial end-to-end latency, and results in slower inference (lower RTFx, where RTFx measures the ratio of audio duration to processing time).
This limitation becomes particularly severe in real-time conversational settings where batch processing is not feasible, as the inability to parallelize token generation directly translates to high per-utterance latency.
Moreover, these systems discard the initial hypothesis often produced by the speech encoder, despite it frequently providing a reasonable draft that could be refined rather than regenerated from scratch.
This work focuses on addressing these limitations and enables parallelizable LLM-based inference.

Connectionist Temporal Classification (CTC) \cite{graves2006connectionist} provides an efficient alternative mechanism for mapping long acoustic sequences to shorter token sequences.
A CTC encoder produces frame-level token posteriors, and inference collapses repeated tokens and removes blanks using an efficient, fully parallel decoding procedure.
However, CTC decoding is constrained by conditional independence and monotonic alignment assumptions.
Moreover, CTC models have limited language modeling capabilities and lack the broad linguistic priors of large pretrained language models, limiting their ability to recover plausible content when acoustic evidence is weak.
In practice, CTC outputs often exhibit local errors such as phonetic substitutions or structural errors such as word deletions, especially in noisy or ambiguous conditions.
Many of these errors are systematic and locally correctable, making CTC hypotheses well-suited as drafts for downstream editing rather than full regeneration.

\begin{figure}[tb]
    \centering
    \includegraphics[width=1\linewidth]{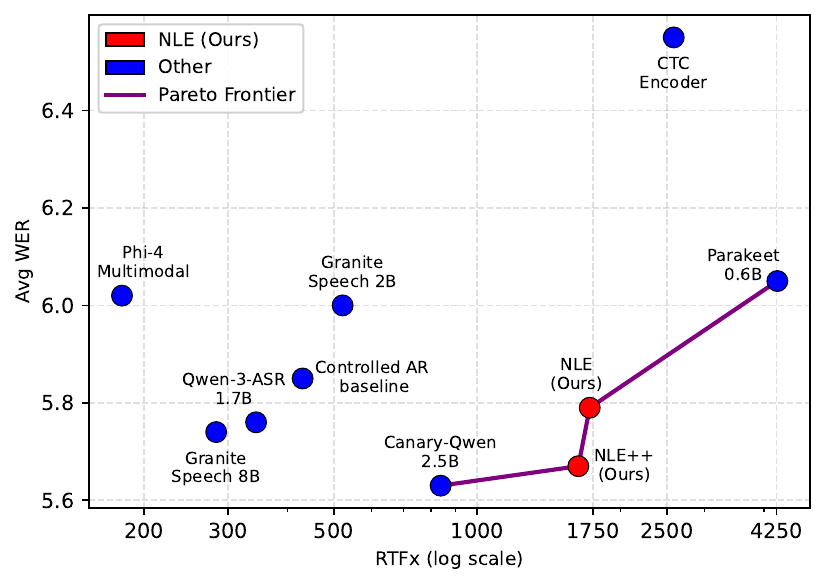}
    \caption{Open ASR leaderboard WER-RTFx tradeoff comparing \method\ and \method++ against top-6 models (as of Feb 2026). Both \method\ variants lie on the Pareto frontier (no other model achieves both lower WER and higher RTFx), achieving competitive accuracy with superior inference speed.}
    \label{fig:pareto}
\end{figure}

We introduce \textbf{\method}, a \textbf{N}on-autoregressive \textbf{L}LM-based \textbf{E}diting ASR system. 
\method\ reframes LLM-based speech recognition as conditional transcript editing: rather than decoding tokens autoregressively, \method\ edits a hypothesis extracted from a pretrained speech encoder, guided by acoustic context from the same encoder.
This editing formulation enables fully parallel prediction, thereby achieving fast inference.

Our approach adapts a pretrained LLM to the editing task, in order to leverage its linguistic knowledge. 
We modify the attention mechanism in the LLM from causal to bidirectional \cite{gong2025scaling,behnamghaderllm2vec}, enabling non-autoregressive editing with full context.
The model is trained using a CTC-style objective with latent alignments \cite{saharia2020non}, which naturally handles the variable-length mapping between input and target transcripts.
We use lightweight LoRA adapters \cite{hu2022lora} to adapt the model to the bidirectional attention and editing objective.
The LoRA adapters can be disabled and the attention mechanism reconfigured back to causal mode, restoring the original LLM functionality.
This design allows the LLM weights to be shared across both ASR and downstream tasks such as question answering on transcribed text -- a key use case for speech-enabled LLMs \cite{saon2025granite}.

We use an interleaved token layout with explicit insertion slots that enables local insertions with minimal token movement.
This design exploits the identity mapping bias of Transformers -- the tendency to copy input tokens unchanged through residual connections and tied embeddings (see Section~\ref{sec:identity_bias}) -- allowing the model to focus on corrections rather than full reconstruction.

On the Open ASR leaderboard \cite{srivastav2025open}, a challenging benchmark containing 39 strong ASR models, \method++ achieves 5.67\% average WER with an RTFx of 1630.
As shown in Figure~\ref{fig:pareto}, both \method\ and \method++ are on the Pareto frontier in the WER-RTFx space, offering a strong accuracy-speed tradeoff compared to leading models.
In single-utterance inference, \method\ achieves 27x speedup over the autoregressive baseline, demonstrating the substantial practical advantages of non-autoregressive decoding for latency-critical applications.

\section{Related Work}
\label{sec:related}
\subsection{LLM-based ASR}
Recent work has explored integrating large language models into ASR systems by conditioning pretrained LLMs on speech representations via learned projection layers \cite{tang2024salmonn,ma2024embarrassingly,chu2024qwen2,saon2025granite}.
Similar to encoder-decoder models like Whisper \cite{radford2023robust}, these approaches pair a speech encoder with a text decoder, but leverage pretrained LLMs to benefit from their linguistic knowledge.
By exploiting the strong linguistic priors of LLMs, they improve transcription accuracy, particularly in challenging acoustic conditions or when handling rare words and proper nouns.
However, most LLM-based ASR systems rely on autoregressive decoding, where tokens are generated sequentially, inherently limiting parallelism and resulting in high inference latency and lower throughput -- a critical bottleneck for real-time applications.
Moreover, autoregressive models can hallucinate plausible but incorrect content when acoustic evidence is weak or ambiguous, and they discard the initial hypothesis often produced by the speech encoder despite it frequently providing a reasonable draft.

Alternative architectures such as RNN-Transducers (RNN-T) \cite{graves2012sequence} and Token-and-Duration Transducers (TDT) \cite{xu2023efficient, rekesh2023fast} offer faster inference through streaming capabilities, but they still generate tokens sequentially and lack access to the broad linguistic knowledge encoded in pretrained LLMs.
Our work addresses these limitations by using an LLM as a \emph{non-autoregressive editor} that refines an initial CTC hypothesis in parallel, combining the speed of NAR decoding with the linguistic knowledge of pretrained LLMs.

\subsection{NAR ASR}
Non-autoregressive (NAR) ASR methods aim to reduce inference latency by predicting tokens in parallel.
Connectionist Temporal Classification (CTC) \cite{graves2006connectionist} is the most widely adopted NAR approach, marginalizing over alignments using dynamic programming and enabling efficient parallel decoding.
However, CTC models typically lack strong language modeling capabilities and are constrained by conditional independence assumptions, limiting their ability to leverage linguistic context for disambiguation~\cite{hannun2017sequence,chan2015listen}.
Several NAR refinement approaches have been proposed to improve upon CTC.
SoftCorrect \cite{leng2023softcorrect} applies constrained CTC loss for transcript correction.
Mask-predict methods \cite{higuchi2020mask,fang2022non,ghazvininejad2019mask} and iterative refinement methods \cite{chan2020imputer,leng2021fastcorrect,chi2021align} perform multiple passes, addressing CTC's conditional independence by conditioning on partial predictions.

Despite offering substantial speed advantages over autoregressive models, NAR methods often struggle to perform insertions (i.e., correcting deletion errors) and to maintain long-range linguistic consistency.
Many rely on fixed-length predictions or masking strategies that make insertions difficult or require multiple refinement iterations.
Our approach builds on CTC's efficiency while addressing these limitations through two key ideas: an insertion-aware interleaved representation that enables local insertions without sequence-wide shifts, and bidirectional LLM-based editing that leverages pretrained linguistic knowledge for improved contextual reasoning.

\subsection{ASR Correction}
Post-processing methods that correct first-pass ASR outputs have been extensively studied.
Traditional approaches include N-best list rescoring and lattice-based rescoring \cite{mangu2000finding,sak2010fly,hu2020deliberation}, which reevaluate hypotheses using refined acoustic or language models.
More recently, LLM-based correction methods have emerged, where ASR transcripts are fed to external LLMs for zero-shot or few-shot correction \cite{ma2023can, ma2025asr}, leveraging their linguistic knowledge.
Supervised error-correction approaches train encoder-decoder models on pairs of erroneous transcripts and references \cite{ma2023n}, enabling learned correction patterns.
Unlike these post-processing methods that operate on finalized ASR outputs, our approach integrates correction into the decoding process itself through non-autoregressive editing conditioned on acoustic embeddings, enabling joint acoustic-linguistic refinement rather than text-only correction.

\subsection{NAR Text Editing and Translation}
Non-autoregressive methods for machine translation and text editing share conceptual similarities with ASR, as both involve mapping between closely related input-output sequences.
Early work \cite{gu2018non} introduced NAR translation using fertility-based parallel generation, while the Levenshtein Transformer \cite{gu2019levenshtein} models editing through explicit operations like KEEP, DELETE, and INSERT.
Most relevant to our work, latent alignment models with CTC have been applied to NAR translation \cite{saharia2020non} and extended to text editing with an explicit COPY operation \cite{zhang2023non}.
We adapt these latent alignment techniques to speech recognition by conditioning on acoustic embeddings rather than text alone, and by leveraging a pretrained causal LLM adapted to bidirectional attention instead of training from scratch.

\begin{figure*}[htb]
    \centering
    \includegraphics[width=0.99\linewidth]{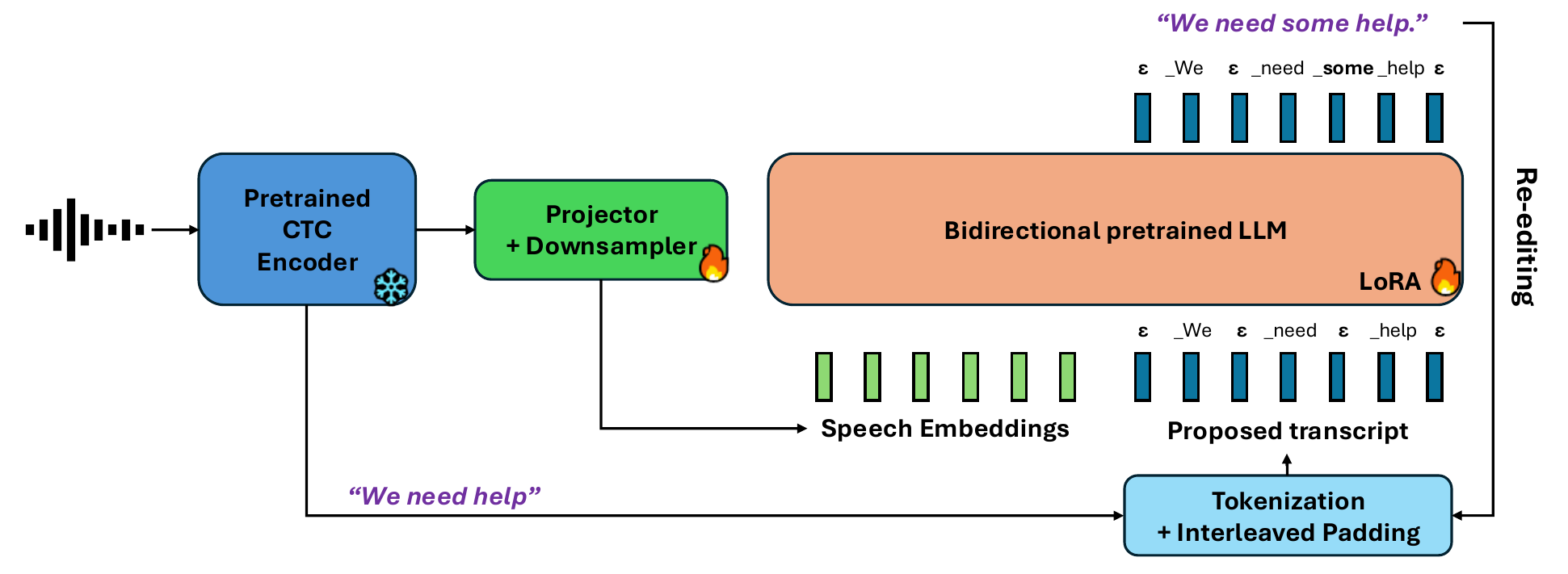}
    \caption{Overview of \method\ architecture. The frozen pretrained CTC encoder produces acoustic embeddings and an initial CTC hypothesis. The hypothesis is tokenized and interleaved with insertion slots ($\epsilon$), then concatenated with the projected speech embeddings. The LoRA-adapted bidirectional LLM editor predicts the edited transcript using a CTC objective. The output can be iteratively re-edited (see Section~\ref{sec:multistep}).}
    \label{fig:overview}
\end{figure*}

\section{Method}
\label{sec:method}
\method\ takes an input utterance and processes it through a pretrained speech encoder to produce acoustic embeddings and a CTC transcript hypothesis.
An LLM-based NAR editor then predicts an edited transcript using a CTC-style objective over a token sequence interleaved with insertion slots.
Figure~\ref{fig:overview} illustrates the complete pipeline.

\subsection{Extracting Hypothesis and Embeddings}
We use a pretrained CTC-based speech encoder which we freeze during training.
We freeze the encoder to preserve its well-trained acoustic modeling capabilities, as the quality of its hypothesis directly affects the entire training process; we leave joint fine-tuning for future work.
The encoder processes the input audio and outputs frame-level embeddings $H \in \mathbb{R}^{T \times d}$, where $T$ is the number of frames and $d$ is the embedding dimension.
It also produces frame-level CTC logits over a character-based vocabulary, which are converted to a character-level string hypothesis using greedy decoding (argmax followed by blank removal and deduplication).
This hypothesis serves as the initial draft for the editing model, while the embeddings $H$ provide acoustic context to guide the refinement process.

\subsection{Retokenization and Interleaved Insertion Slots}
The character-level CTC hypothesis is re-tokenized using the LLM's subword tokenizer to align with the LLM's vocabulary.
Since the LLM tokenizer is trained on well-formed text, misspelled words in the CTC hypothesis (e.g., "philossophy") may not have a corresponding token and will be split into multiple subword units.
We denote the resulting token sequence as $x = (x_1, \ldots, x_N)$.

To efficiently handle insertion edits, we construct an interleaved sequence with explicit insertion slots:
\begin{equation}
\tilde{x} = (\epsilon, x_1, \epsilon, x_2, \ldots, \epsilon, x_N, \epsilon).
\label{eq:interleaved_padding}
\end{equation}
where $\epsilon$ denotes a blank symbol (we reuse the LLM's EOS token).
This interleaved representation creates $N+1$ insertion slots (one before each token, and one after the last token), enabling local insertions without requiring the model to shift the entire downstream sequence.
Single-token insertions can be handled by filling one insertion slot.
Multi-token insertions require shifting only the neighboring tokens within a local block to accommodate the new content, while the rest of the sequence remains unaffected.

To illustrate this, consider adding the tokens $(a,b,c)$ between $x_k$ and $x_{k+1}$. 
Given the original sequence:
$$(\dots \epsilon, x_k, \epsilon, x_{k+1}, \epsilon, \dots)$$
the model can produce this insertion by making the following per-position predictions: moving $x_k$/$x_{k+1}$ to the left/right $\epsilon$ spot, and using the 3 freed-up middle positions to add $(a,b,c)$:
$$(\dots, x_k, a, b, c, x_{k+1}, \dots)$$
This can be performed within a single forward pass, keeping the rest of the tokens intact.
An insertion of $K$ tokens requires changing $2K-1$ tokens from the original sequence, while keeping the remaining tokens unchanged.
The maximum number of tokens that can be inserted is $N+1$, which was never required in our training data\footnote{We set the minimum number of input tokens to 8 to allow sufficient insertion capacity for short utterances.}.

\subsection{Identity Mapping Bias in Transformers}
\label{sec:identity_bias}
Our editing approach exploits the identity mapping bias inherent to Transformer architectures \cite{he2016deep, he2016identity}.
This bias arises from two key architectural properties:
\begin{itemize}
    \item \textbf{Residual connections} preserve input representations across layers, allowing information to flow unchanged through the network.
    \item \textbf{Tied input-output embeddings} are used in most LLMs (including Granite 4.0 1B Base we used in \method) and create a strong copying bias: the dot product between a token's embedding and itself in the output projection matrix tends to be high, making the model naturally inclined to predict the input token.
\end{itemize}
When the interleaved sequence is decoded without edits, it naturally recovers the original hypothesis.
This property is beneficial for our task, as most tokens in the CTC hypothesis are correct and should be preserved.
The model can focus its learning capacity on identifying and correcting errors rather than reconstructing the entire transcript, which is crucial for maintaining strong performance while enabling fast parallel decoding.
We further reinforce this bias through a copying regularization objective (Section~\ref{sec:cr_loss}).
In Figure~\ref{fig:overview}, for example, the model is only required to change $x_5=\epsilon$ to "\_some", while copying the rest of the input tokens, with no token displacements needed due to the interleaved insertion slots. 

\subsection{Bidirectional LLM-based Editor}
We compute acoustic embeddings using a learned projector $P_\theta$ that downsamples the frame-level speech embeddings $H$ and maps them into the LLM's embedding space.
The interleaved hypothesis $\tilde{x}$ is embedded using the LLM's token embedding table $E$.
These representations are concatenated along the sequence dimension to form the input to the LLM:
$Z = [P_\theta(H); E(\tilde{x})] \in \mathbb{R}^{(T' + 2N+1) \times d_{\text{LLM}}}$,
where $T'$ is the downsampled acoustic sequence length and $d_{\text{LLM}}$ is the LLM's hidden dimension.

Starting from a pretrained causal LLM, we modify its attention mechanism to bidirectional by removing the causal attention mask, allowing each position to attend to all other positions, while keeping the positional embeddings unchanged.
This bidirectional context is essential for effective editing, as corrections often require information from both past and future tokens.
We adapt the LLM using LoRA (Low-Rank Adaptation), which enables efficient fine-tuning while preserving the model's pretrained linguistic knowledge.
The LoRA adapters can be disabled to restore the original causal LLM functionality, allowing the same model weights to be shared across ASR and other downstream tasks.

\subsection{CTC-based Editing Objective}
The LLM editor outputs token logits $L \in \mathbb{R}^{(2N+1)\times|\mathcal{V}|}$ over the vocabulary $\mathcal{V}$ for each position in the interleaved sequence $\tilde{x}$.
We apply the standard CTC loss $\mathcal{L}_{\mathrm{CTC}}(L, x^\star)$ that marginalizes over all valid alignments between the predicted logits and the reference transcript $x^\star$ using dynamic programming.
The CTC objective naturally handles the variable-length mapping between the interleaved input sequence and the target transcript, allowing the model to learn which positions to keep, delete, or use for insertions.
This design choice avoids the need to pre-align the reference and hypothesis, as alignment happens implicitly during the loss calculation through CTC's dynamic programming.

Edit operations are performed as follows:
\begin{itemize}
    \item \textbf{Copy:} The identity bias enables copying tokens from the input hypothesis by default.
    \item \textbf{Replace:} Substitutions are performed by predicting a different token at a position.
    \item \textbf{Delete:} Deletions are performed by predicting the blank symbol $\epsilon$ at a token position.
    \item \textbf{Insert:} Single-token insertions use the explicit insertion slots. Multi-token insertions require shifting neighboring tokens, which the model learns through the CTC loss that allows multiple valid alignments.
\end{itemize}

\subsection{Copying Regularization Objective}
\label{sec:cr_loss}
While the CTC objective allows the model to learn editing operations, it does not explicitly enforce the identity mapping bias.
The Transformer layers mix features across positions, and the CTC loss permits multiple alignments that produce the same output text, including redundant edits (changes in the CTC lattice that collapse to the same decoded string).
When adapting a pretrained LLM trained with next-token prediction, this bias may be overridden by the model's learned next-token behavior.
To encourage the model to preserve correct tokens and make edits more interpretable, we add an auxiliary copying regularization (CR) loss:
\begin{equation}
\mathcal{L}_{\mathrm{CR}} = -\sum_{i=1}^{2N+1} \log P(\tilde{x}_i | L_i),
\end{equation}
where $L_i$ are the logits at position $i$ and $\tilde{x}_i$ are the input tokens from $\tilde{x}$.
This cross-entropy loss encourages the model to predict the input tokens at each position, reinforcing the copying bias.
The total training objective becomes:
\begin{equation}
\mathcal{L} = \mathcal{L}_{\mathrm{CTC}} + \lambda \mathcal{L}_{\mathrm{CR}},
\end{equation}
where $\lambda$ controls the regularization strength.
Since the CR loss actively encourages predicting the input sequence exactly, it is important to keep $\lambda$ much smaller than the CTC loss to avoid suppressing necessary edits.

\subsection{Inference}
At inference, we extract acoustic embeddings $H$ and the CTC hypothesis from the frozen speech encoder.
We retokenize the hypothesis and interleave it with blank symbols to form $\tilde{x}$, which is then concatenated with the projected acoustic embeddings and fed to the bidirectional LLM editor.
The editor produces output logits $L$ for each position in the interleaved sequence.
We apply CTC greedy decoding (argmax followed by blank removal and deduplication) to obtain the final edited transcript $\hat{x}$.

Crucially, all predictions are made in parallel across the sequence, enabling significantly faster inference compared to autoregressive decoding.
The computational cost is dominated by the single forward pass through the LLM, which processes all positions simultaneously.
Optionally, the editing process can be iterated by feeding $\hat{x}$ back as input for additional refinement steps, though this comes at the cost of additional forward passes (see Sec.~\ref{sec:multistep} for analysis).

\section{Experiments}
\label{sec:exp}
\subsection{Training}

\textbf{Model Architecture.}
The CTC encoder contains 440M parameters and is based on a 16-layer Conformer architecture \cite{gulati2020conformer} with block attention using a block size of 200 frames (corresponding to 4 seconds at a 50 Hz frame rate).
The encoder operates on 16 kHz audio with stacked log-mel features (80 mel bins, 2-frame stacking) and self-conditioning at layer 8 \cite{saon2025granite}.
The CTC vocabulary consists of 384 character-based units.
We extract hidden layer representations from four intermediate layers of the CTC encoder (layers 4, 8, 12, and 16), which are concatenated along the feature dimension to provide multi-scale acoustic representations.

To map the acoustic representations, we train a 1-layer Q-Former projector \cite{li2023blip} that downsamples the concatenated hidden layers by 5x (from 15-frame windows to 3 queries).
Each query consists of the mean-pooled representation of its corresponding 5-frame segment and cross-attends to the full 15-frame window with learnable positional embeddings.
The Q-Former has hidden dimension 1024, feedforward dimension 2048, and 16 attention heads with head dimension 64.
Our base language model is the lightweight Granite 4.0 1B base \cite{granite2025}, which we adapt using LoRA with rank 128, applied to both attention and MLP layers.
The total number of trainable parameters (projector + LoRA) is only 14M.
For the CTC blank symbol $\epsilon$, we reuse the EOS token from the LLM vocabulary to avoid modifying the vocabulary (any special token would suffice).

\textbf{Training Procedure.}
Models are trained for 3 epochs (180K steps total) using the AdamW optimizer, with a peak learning rate of 3e-5 and a cosine schedule with 5\% warmup and minimum learning rate of 1\% of peak.
We adopt the balanced sampling strategy of \cite{saon2025granite} where the $\alpha\in[0,1]$ hyperparameter controls the dataset balancing level and set $\alpha=0.65$ (compared to $\alpha=0.6$ in \cite{saon2025granite}) for all experiments.
The copying regularization weight is set to $\lambda=0.02$ (a small value to avoid dominating the CTC loss).
We use a global batch size of approximately 320 utterances, spread across 8 H100 GPUs.
Training takes approximately 30 hours for 3 epochs on 8 H100 GPUs, compared to 24 hours for the AR baseline (1.25x longer due to the interleaved padding increasing sequence length).

\subsection{Datasets}
Our training datasets include AMI (IHM+SDM) \cite{kraaij2005ami}, VoxPopuli \cite{wang2021voxpopuli}, YODAS (9K hours subset) \cite{li2023yodas}, CommonVoice 15 \cite{ardila2020common}, MLS \cite{pratap2020mls}, Earnings22 \cite{del2022earnings}, Fisher \cite{cieri2004fisher}, CallHome \cite{canavan1997callhome}, and SwitchBoard \cite{godfrey1992switchboard}.
We train on five languages: English, Spanish, French, German, and Portuguese.
The combined training corpus comprises approximately 70K hours of speech across five languages.
For evaluation, we use the official validation and test sets of the training datasets, as well as three additional test-only datasets: GigaSpeech \cite{chen2021gigaspeech}, TED-LIUM \cite{rousseau2012ted}, and SPGISpeech \cite{o2021spgispeech}.

During training, we filter out utterances longer than 80 seconds and examples where the character-to-frame ratio exceeds 0.66.
We apply SpecAugment (time and frequency masking) and noise augmentation with 25\% probability using MUSAN and FreeSound backgrounds at SNR levels between -5 and 20 dB.

\subsection{Evaluation Protocol}
We evaluate \method\ in terms of transcription accuracy and inference efficiency, focusing on the WER-RTFx trade-off.
All RTFx measurements are obtained using offline batched inference on a single H100 GPU with bf16 precision.
For our AR and NAR models, we use a batch size of 96 utterances.
To assess performance in latency-critical conversational settings, we additionally measure RTFx with batch size 1, where the inability to parallelize computations in autoregressive models becomes more severe.
For other models on the Open ASR leaderboard, we use the open-source optimized implementations from the leaderboard repository, measured on the same H100 hardware to ensure fair comparison.
We use CTC greedy decoding (argmax+collapse) for CTC-based models, as beam search yields marginal improvements at significantly higher inference cost (we observed marginal WER improvement with 5-10x slower RTFx).
For autoregressive models, we use greedy autoregressive decoding.
WER is computed in lowercase, applying Whisper normalization and using the jiwer library.
We report several aggregated results: Open ASR leaderboard \cite{srivastav2025open} is the average WER across the datasets included in the leaderboard; the CV and MLS metrics are averages across all reported CommonVoice 15 and Multilingual LibriSpeech subsets, respectively.

\begin{table}[tb]
\centering
\caption{WER (\%, lower is better) and RTFx (higher is better) comparison. \method\ competes with AR accuracy while being 4$\times$/27$\times$ faster in batched/single-utterance inference.}
\label{tab:results}
\begin{tabular}{lccc}
\toprule
\textbf{Dataset} & \textbf{NLE} & \textbf{AR} & \textbf{CTC} \\
\midrule
\textsc{ami-ihm} & 8.3 & 8.6 & 9.4 \\
\textsc{ami-sdm} & 21.4 & 23.8 & 24.4 \\
\textsc{cv15-de} & 5.6 & 4.7 & 6.3 \\
\textsc{cv15-en} & 7.3 & 7.1 & 9.5 \\
\textsc{cv15-es} & 5.0 & 4.1 & 5.5 \\
\textsc{cv15-fr} & 8.2 & 7.2 & 10.8 \\
\textsc{cv15-pt} & 3.0 & 2.7 & 3.4 \\
\textsc{earnings} & 10.0 & 10.1 & 11.5 \\
\textsc{gigaspeech} & 10.1 & 10.0 & 10.6 \\
\textsc{ls-clean} & 1.4 & 1.5 & 1.7 \\
\textsc{ls-other} & 3.1 & 3.1 & 3.7 \\
\textsc{mls-de} & 4.7 & 4.5 & 4.9 \\
\textsc{mls-en} & 4.8 & 4.7 & 5.7 \\
\textsc{mls-es} & 3.5 & 3.1 & 3.7 \\
\textsc{mls-fr} & 4.6 & 4.5 & 5.6 \\
\textsc{mls-pt} & 10.0 & 10.1 & 8.5 \\
\textsc{spgi} & 3.5 & 3.5 & 4.5 \\
\textsc{ted-lium} & 3.9 & 3.7 & 3.9 \\
\textsc{vox} & 6.2 & 6.2 & 7.1 \\
\midrule
\addlinespace[0.3em]
\multicolumn{4}{l}{\textit{Aggregate Metrics}} \\
\rowcolor{blue!10}Average (All 19) & 6.54 & 6.48 & 7.40 \\
\rowcolor{blue!10}Open ASR Average & 5.79 & 5.82 & 6.55 \\
\rowcolor{blue!10}CV Average & 5.79 & 5.18 & 7.10 \\
\rowcolor{blue!10}MLS Average & 5.51 & 5.39 & 5.66 \\
\addlinespace[0.3em]
\midrule
\multicolumn{4}{l}{\textit{Speed Metrics}} \\
\rowcolor{green!10}RTFx: Batch size 96 & 1722 & 430 & 2584 \\
\rowcolor{green!10}RTFx: Batch size 1 & 322 & 12 & 760 \\
\bottomrule
\end{tabular}
\end{table}

\section{Results}
\label{sec:results}
\subsection{Open ASR Leaderboard Comparison}
We compare \method\ against the top-6 models on the Open ASR leaderboard at the time of submission: Canary-Qwen 2.5B \cite{sekoyan2025canary}, Granite Speech 2B and 8B \cite{saon2025granite}, Phi-4 Multimodal \cite{abouelenin2025phi}, Qwen3-ASR 1.7B \cite{shi2026qwen3}, and Parakeet 0.6B \cite{sekoyan2025canary}.
Figure~\ref{fig:pareto} shows that \method\ is on the Pareto frontier, ranking 4th in average WER (5.79\%) while achieving 1722 RTFx.
Among the top-6 models, only Parakeet (4264 RTFx) is faster, but at the cost of higher WER (6.05\%).
\method\ demonstrates a superior accuracy-speed tradeoff compared to other LLM-based systems, achieving competitive accuracy with 2-10x faster inference than models like Canary-Qwen, Granite Speech, Qwen3-ASR, and Phi-4 Multimodal.
Notably, \method\ is the only model on the Pareto frontier on the Open ASR leaderboard that is multilingual, supporting five languages (English, Spanish, French, German, and Portuguese), while Canary-Qwen and Parakeet are English-only models.

\subsubsection{NLE++ (Enhanced Training)}
\label{sec:nlepp}
We additionally trained \textbf{\method++}, a variant with the same architecture and data but with several training improvements: a larger projector (2 layers, 2$\times$ hidden and feedforward dimensions) with input/output dropout of 0.1, LoRA rank increased from 128 to 160, a 2$\times$ higher peak learning rate (6e-5), 5 training epochs (vs.\ 3), a 2$\times$ larger global batch size (640 utterances across 16 H100 GPUs on 2 nodes), and maximum audio length extended to 120 seconds (vs.\ 80 seconds).
The larger projector and LoRA rank scale the total number of trainable parameters to 280M (160M projector + 120M LoRA), compared to 14M in \method.
As \method++ differs in training budget and configuration from the controlled setup in Section~\ref{sec:results}, we report it separately in Table~\ref{tab:results_pp}.
\method++ achieves 5.67\% Open ASR WER -- a 0.12\% absolute improvement over \method\ and improves the all-19 average from 6.54\% to 6.44\%.
RTFx decreases modestly to 1630 (vs. 1722) due to the larger projector, but \method++ remains on the Pareto frontier (Figure~\ref{fig:pareto}).
These results suggest that \method\ benefits from scaling up training compute and model capacity, pointing toward further improvements with additional resources.

\begin{table}[tb]
\centering
\caption{Aggregate WER (\%) and RTFx comparison of \method\ vs.\ \method++.}
\label{tab:results_pp}
\begin{tabular}{lcc}
\toprule
& \textbf{NLE} & \textbf{NLE++} \\
\midrule
\multicolumn{3}{l}{\textit{Aggregate Metrics}} \\
\rowcolor{blue!10}Average (All 19) & 6.54 & 6.44 \\
\rowcolor{blue!10}Open ASR Average & 5.79 & 5.67 \\
\rowcolor{blue!10}CV Average & 5.79 & 5.41 \\
\rowcolor{blue!10}MLS Average & 5.51 & 5.76 \\
\addlinespace[0.3em]
\midrule
\multicolumn{3}{l}{\textit{Speed}} \\
\rowcolor{green!10}RTFx: Batch size 96 & 1722 & 1630 \\
\rowcolor{green!10}RTFx: Batch size 1 & 322 & 310 \\
\bottomrule
\end{tabular}
\end{table}

\subsection{Controlled Evaluation}
We conduct a controlled comparison against two baselines using the same training setup:
\begin{itemize}
    \item \textbf{CTC encoder only:} Greedy decoding from the pretrained speech encoder.
    \item \textbf{Controlled AR baseline:}
    To enable a fair and controlled comparison between AR and NAR approaches, we train an AR LLM-based ASR system using the same encoder, projector, LLM backbone, datasets, balanced sampling strategy, and optimization setup described above.
    The systems differ only in the decoding strategy: autoregressive next-token prediction versus non-autoregressive editing.
    The AR baseline projects acoustic embeddings using the same projector, then concatenates them with the ground-truth transcript tokens wrapped with BOS and EOS tokens from the LLM vocabulary.
    The model is trained using standard next-token prediction with cross-entropy loss in causal attention mode.
    At inference, the model generates transcripts autoregressively using greedy decoding.
\end{itemize}

Table~\ref{tab:results} presents results from our controlled evaluation across 19 test datasets covering 5 languages.
\method\ consistently outperforms the CTC encoder baseline, reducing average WER from 7.40\% to 6.54\% while maintaining high inference speed.
\method\ consistently improves over CTC, outperforming it on 17 out of 19 test sets.
Compared to the controlled AR baseline, \method\ achieves comparable accuracy with significantly faster inference: 4x speedup in batched scenarios (1722 vs 430 RTFx) and 27x speedup in single-utterance settings (322 vs 12 RTFx).
The latency advantage is particularly pronounced in real-time conversational applications where batching is not possible, as the autoregressive model's inability to parallelize becomes a critical bottleneck.
We note that \method\ underperforms the AR baseline on CommonVoice subsets, showing slightly worse performance in multilingual settings, possibly because the CTC encoder's training data is predominantly English, resulting in weaker non-English hypotheses that \method\ must then correct, compounded by the LLM tokenizer's English-centric BPE vocabulary.
These results confirm the efficacy of our proposed approach.

\subsection{Ablation Study}
\label{sec:ablation}
\begin{figure}[tb]
    \centering
    \includegraphics[width=0.98\linewidth]{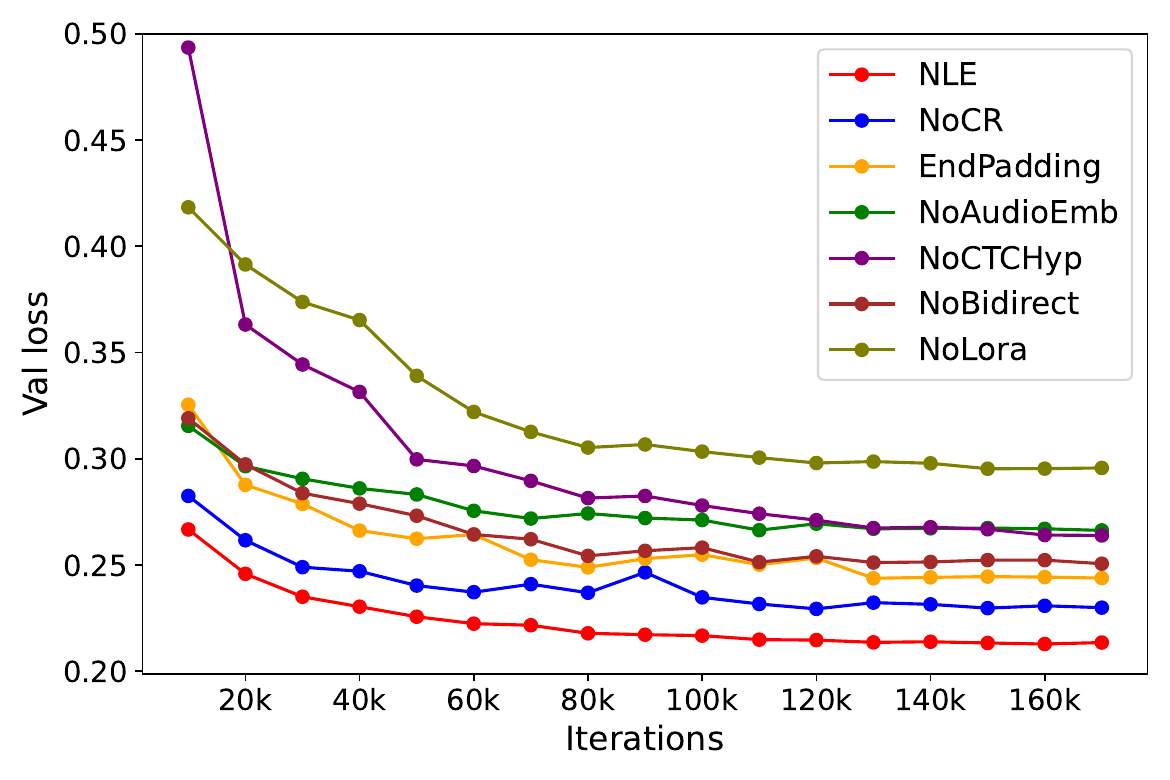}
    \caption{Validation loss over training steps for ablation study (see Section~\ref{sec:ablation}). \method\ (full model) achieves the lowest validation loss, confirming that each design choice contributes positively to overall performance.}
    \label{fig:ablation}
\end{figure}

Figure~\ref{fig:ablation} analyzes the impact of key design choices on validation loss, which consistently correlates with better WER on test sets.

\textbf{Copying regularization (CR) loss.}
Removing the CR loss (NoCR) degrades validation loss, even though the reported loss includes the non-negative CR term, meaning the full model still reaches a lower loss despite the added penalty.
This demonstrates that CR improves both training stability and final performance by explicitly encouraging the identity mapping bias.

\textbf{Bidirectional attention.}
Restricting the LLM to causal attention (NoBidirect) limits its ability to leverage future context for editing.
Notably, enabling bidirectional attention consistently improves convergence and validation loss, despite the LLM being pretrained with causal attention, confirming that future context is critical for effective non-autoregressive editing.

\textbf{Padding strategy.}
Interleaved padding significantly outperforms end-of-sequence padding (EndPadding).
End padding causes large token displacements during insertions, while interleaved padding preserves token locality, aligning with the Transformer's locality bias.

\textbf{Audio conditioning.}
Removing acoustic embeddings (NoAudioEmb; multiplying them by 0) substantially degrades performance, highlighting the necessity of acoustic grounding for accurate hypothesis refinement.

\textbf{Hypothesis Conditioning.}
Removing the input hypothesis as conditioning (NoCTCHyp; replacing it with a sequence of blanks with the same length) also degrades performance, suggesting the model struggles with predicting the entire hypothesis from scratch.

\textbf{LoRA adaptation.}
Keeping the LLM frozen (NoLoRA) degrades validation loss, highlighting that LoRA adaptation is important for optimal performance.

\subsection{Blank Density}
\label{sec:blank_density}
We investigate the impact of varying the density of insertion slots in the interleaved sequence.
Instead of inserting a blank between every token (Every 1), we experiment with inserting blanks every 2/3 tokens.
Table~\ref{tab:blank_density} shows that reducing blank density degrades accuracy while providing minimal speedup, as the sequence length is dominated by acoustic tokens rather than text tokens.
Inserting blanks between every token (Every 1) provides the best accuracy-speed tradeoff.
\begin{table}[tb]
\centering
\caption{Impact of blank density (adding an insertion slot every K tokens) on the average WER and RTFx. Inserting a blank between every token (Every 1) achieves superior WER with negligible RTFx drop compared to sparser insertion strategies.}
\label{tab:blank_density}
\begin{tabular}{lcc}
\toprule
\textbf{Blank Density} & \textbf{Average WER (\%)} & \textbf{RTFx} \\
\midrule
Every 1 (NLE) & 6.54 & 1722 \\
Every 2 & 6.80 & 1750 \\
Every 3 & 6.91 & 1770 \\
\bottomrule
\end{tabular}
\end{table}

\subsection{Multi-Step Editing}
\label{sec:multistep}
We investigate whether iteratively applying the editor can further improve transcription accuracy.
In multi-step editing, the editor's output logits are decoded using CTC (argmax + collapse) and re-interleaved with insertion slots before being fed back as input to the editor for another round of refinement, while keeping the acoustic embeddings fixed.
This process can be repeated multiple times, with each step potentially correcting errors introduced or missed in previous iterations.
Table~\ref{tab:multistep} suggests that applying a second editing step yields modest improvements, at the cost of reduced inference speed.
However, further iterations show diminishing returns, with the third step slightly degrading performance.
This degradation can be attributed to distribution mismatch: the editor is trained on CTC hypotheses, but at inference iteratively processes its own outputs, which have different error characteristics.
Text augmentation strategies during training could potentially address this mismatch and improve multi-step refinement.
For most applications, single-step editing provides the best accuracy-speed tradeoff.

\begin{table}[tb]
\centering
\caption{Average WER and RTFx for multi-step editing. Step 0 corresponds to the CTC baseline (no editing). A 2-step edit yields negligible accuracy gain at the cost of reduced RTFx, while a 3-step edit degrades performance below the single-step result.}
\label{tab:multistep}
\begin{tabular}{lcc}
\toprule
\textbf{Editing Steps} & \textbf{Average WER (\%)} & \textbf{RTFx} \\
\midrule
0 (CTC only) & 7.40 & 2584 \\
1 & 6.54 & 1722 \\
2 & 6.53 & 1259 \\
3 & 6.59 & 1082 \\
\bottomrule
\end{tabular}
\end{table}

\begin{figure}[tb]
\centering
\includegraphics[width=0.98\linewidth]{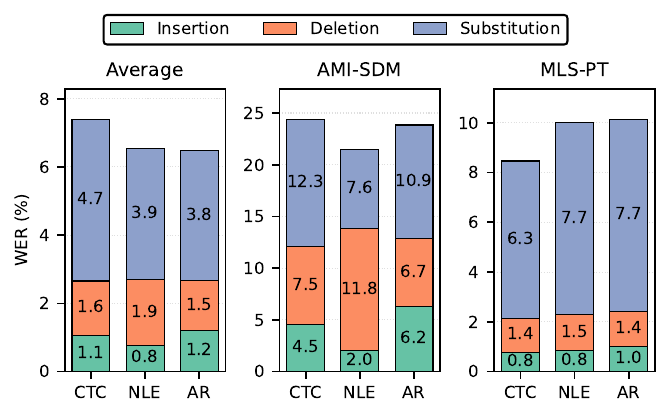}
\caption{Insertion, deletion and substitution rates (\%) for three conditions: average across all datasets, AMI-SDM, and MLS-PT.}
\label{fig:decomposition}
\end{figure}

\subsection{Error Analysis}
\label{sec:error_analysis}
We analyze the types of errors made by different models to understand their behavior.
Figure~\ref{fig:decomposition} shows the decomposition of WER into insertions, deletions, and substitutions for CTC, \method, and AR models, averaged across all datasets as well as for two test-cases: AMI-SDM and MLS-PT.
On average, the AR model exhibits the highest insertion rate, suggesting potential hallucinations when acoustic evidence is weak.
The challenging AMI-SDM dataset illustrates this pattern more dramatically, with AR producing significantly more insertions.
In contrast, \method\ shows the highest deletion rate and lowest insertion rate, reflecting a more conservative editing strategy that prefers deletions over insertions.

Interestingly, on MLS-PT, both \method\ and AR fail to improve over CTC, with \method\ showing particularly high substitution errors.
The MLS-PT training set is very small, making this dataset particularly challenging and potentially leading to overfitting or insufficient adaptation to Portuguese.
This highlights that when the CTC hypothesis quality is relatively good, the editing approach may introduce errors rather than corrections, emphasizing the importance of the initial hypothesis quality.

\begin{table}[tb]
\centering
\caption{Editing examples showing CTC outputs and \method\ corrections. Errors are highlighted in \red{red}, corrections in \blue{blue}, deletions of words in the reference are colored in \orange{orange}.}
\label{tab:examples}
\small
\begin{tabular}{p{0.98\linewidth}}
\toprule
\textbf{Example 1} \\
\textbf{REF:} this thing we are going to design \orange{is} a new remote control \\
\textbf{CTC:} \red{thes ting weare ona design ish} a new remote control \\
\textbf{\method:} \blue{this thing we are going to design} a new remote control \\
\midrule
\textbf{Example 2} \\
\textbf{REF:} often prevented from going to school they are forced to work eighteen to twenty hours per day \\
\textbf{CTC:} often prevented from going to school they are forced \red{set} to work eighteen \red{o} twenty hours per day \\
\textbf{\method:} often prevented from going to school they are forced to work eighteen \blue{to} twenty hours per day \\
\midrule
\textbf{Example 3} \\
\textbf{REF:} both are extremely important but innovation needs to get an extra push \\
\textbf{CTC:} \red{i ar h} extremely important but innovation needs to get an extra push \\
\textbf{\method:} \red{these} \blue{are} extremely important but innovation needs to get an extra push \\
\midrule
\textbf{Example 4} \\
\textbf{REF:} er wurde auf anderen kontinenten ubrigens bereits als wichtiger trendsetter gefeiert \\
\textbf{CTC:} er wurde auf anderen kontinenten ubrigens \red{bereitas} wichtiger \red{trensitte} gefeiert \\
\textbf{\method:} er wurde auf anderen kontinenten ubrigens \blue{bereits als} \red{wichtige trente} gefeiert \\
\midrule
\textbf{Example 5} \\
\textbf{REF:} oreille quand noé eut rassemblé les animaux devant l'arche il se dit \\
\textbf{CTC:} \red{chapitre xvi te ab c petit comte d jules maitre enregistrer pour l'ybrevoxe poir orge} oreille quand \red{noée} eut rassemblé les animaux devant l'arche il se dit \\
\textbf{\method:} \red{oreilles} quand \blue{noé} eut rassemblé les animaux devant l'arche il se dit \\
\bottomrule
\end{tabular}
\end{table}

\subsection{Qualitative Analysis}
\label{sec:qualitative}
To illustrate the types of corrections performed by \method, Table~\ref{tab:examples} presents representative examples comparing CTC outputs with \method\ predictions.
The examples demonstrate how \method\ corrects various error types: incorrect writing of several words (Example 1), substitution corrections (Example 2), and completing plausible content from severely corrupted input (Example 3).
Examples 4-5 showcase multilingual capabilities, with \method\ successfully correcting German and French transcripts.
Notably, in the French example, \method\ removes a long preamble (book title, chapter, and narrator information) that was spoken but should not be transcribed according to the ground truth labels, demonstrating that the model has learned to filter out metadata content based on the training data conventions.
These examples highlight the model's ability to leverage both acoustic embeddings and linguistic priors from the pretrained LLM to produce more accurate transcripts across multiple languages.

\begin{figure}[tb]
    \centering
    \includegraphics[width=0.98\linewidth]{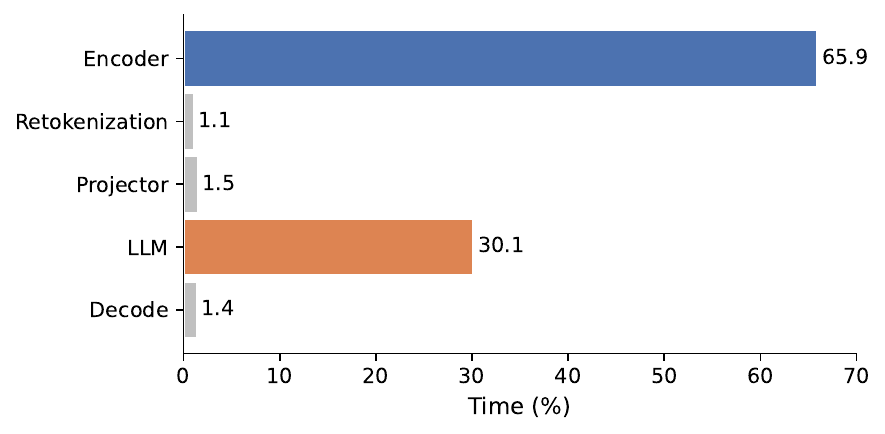}
    \caption{Inference time breakdown across the different stages of \method. The encoder dominates at 66\% of total time, with the LLM contributing $\sim$30\%, and all remaining stages under 4\%.}
    \label{fig:rtfx_breakdown}
\end{figure}

\subsection{Inference Time Breakdown}
Figure~\ref{fig:rtfx_breakdown} shows the distribution of inference time across different stages of \method.
The encoder forward pass dominates at 66\% of total time, accounting for roughly 2$\times$ more compute than the LLM.
We note that the LLM operates on a 5$\times$ shorter sequence relative to the encoder output (10Hz after downsampling vs.\ 50Hz), which explains this result.
The remaining stages -- retokenization, projection, and output decoding -- together account for only 4\% of inference time.
In contrast, an autoregressive system would see the LLM dominate due to sequential token generation, whereas \method's parallel decoding keeps the LLM contribution modest.

\section{Discussion}
\label{sec:discussion}
We introduced \method, a non-autoregressive LLM-based ASR system that reframes speech recognition as conditional transcript editing.
\method\ leverages a pretrained CTC encoder to produce both acoustic embeddings and an initial hypothesis, then applies a bidirectional LLM editor with an interleaved insertion slot representation.
\method\ achieves accuracy comparable to autoregressive baselines while delivering up to 4x faster inference in batched scenarios, with speedup advantages increasing further in single-utterance latency-critical conversational settings.
Our approach demonstrates that pretrained LLMs can be effectively adapted for non-autoregressive editing through lightweight LoRA adapters and attention mask modifications.
The interleaved insertion slot scheme enables efficient handling of insertions while preserving the identity mapping bias of Transformer architectures.

\subsection{Limitations}
\method\ is less flexible than autoregressive models in handling tasks where the output significantly diverges from the input hypothesis.
While \method\ excels at correcting local errors, it is less likely to generalize to tasks requiring substantial changes to the hypothesis, such as spoken question answering, where the expected response is vastly different from the transcript.
Moreover, when the CTC encoder and the LLM use different tokenizers, \method\ requires transferring the hypothesis from GPU to CPU for retokenization and back, adding minor latency overhead.
While using the LLM's tokenizer for CTC training would avoid this overhead, it would significantly increase encoder training costs.

\subsection{Future Work}
Several promising directions could further improve \method's capabilities.
Text augmentation strategies during training could address the distribution mismatch observed in multi-step editing (Section~\ref{sec:multistep}), potentially enabling more effective iterative refinement.
Such augmentations could include synthetic errors, paraphrasing, or using the model's own predictions as training inputs to better match inference-time conditions.
Combining the editing approach with mask-predict strategies could further improve transcription accuracy by allowing the model to iteratively refine uncertain predictions while maintaining parallel inference.
This hybrid approach could leverage the strengths of both paradigms: the efficiency of single-pass editing for most tokens and the refinement capability of masking for challenging regions.

An additional promising direction is to restructure the LLM architecture to process audio and text in separate layers with cross-attention between modalities, reducing computational complexity from quadratic to linear with respect to audio length.
This would be particularly beneficial for long-form audio processing.

Other avenues for future research include:
\begin{itemize} 
    \item Investigating the use of the underlying LLM as a language model for CTC beam decoding to provide stronger linguistic priors during initial hypothesis generation.
    \item Joint fine-tuning of the encoder and editor for end-to-end optimization (though this would require careful training strategies to maintain acoustic modeling performance).
    \item Extending the approach to streaming scenarios through mechanisms that handle partial hypotheses and incomplete acoustic context or chunk-based processing strategies that maintain parallel inference advantages within each chunk.
\end{itemize}

\bibliographystyle{IEEEtran}
\bibliography{mybib}

\end{document}